\documentclass{osa-article}

\journal{osajournal}

\articletype{Research Article}

\usepackage{lineno}
\usepackage{cleveref}
\usepackage{ulem}

\newcommand{\red}[1]{{#1}}

\begin{document}

\title{Two-Photon Interference LiDAR Imaging}

\author{Robbie Murray,\authormark{1} and Ashley Lyons \authormark{1,*}}

\address{\authormark{1}School of Physics and Astronomy, University of Glasgow, Glasgow, G12 8QQ, UK}

\email{\authormark{*}ashley.lyons@glasgow.ac.uk} 



\begin{abstract}
Optical Coherence Tomography (OCT) is a key 3D imaging technology that provides micron scale depth resolution for bio-imaging. This resolution substantially surpasses what it typically achieved in Light Detection and Ranging (LiDAR) which is often limited to the millimetre scale due to the impulse response of the detection electronics. However, the lack of coherence in LiDAR scenes, arising from mechanical motion for example, make OCT practically infeasible. Here we present a quantum interference inspired approach to LiDAR which achieves OCT depth resolutions without the need for high levels of stability. We demonstrate depth imaging capabilities with an effective impulse response of 70 $\mu$m, thereby allowing ranging and multiple reflections to be discerned with much higher resolution than conventional LiDAR approaches. This enhanced resolution opens up avenues for LiDAR in 3D facial recognition, and small feature detection/tracking as well as enhancing the capabilities of more complex time-of-flight methods such as imaging through obscurants and non-line-of-sight imaging.
\end{abstract}

\section{Introduction}
\begin{figure}[t]
\centering
\includegraphics[width=0.8\textwidth]{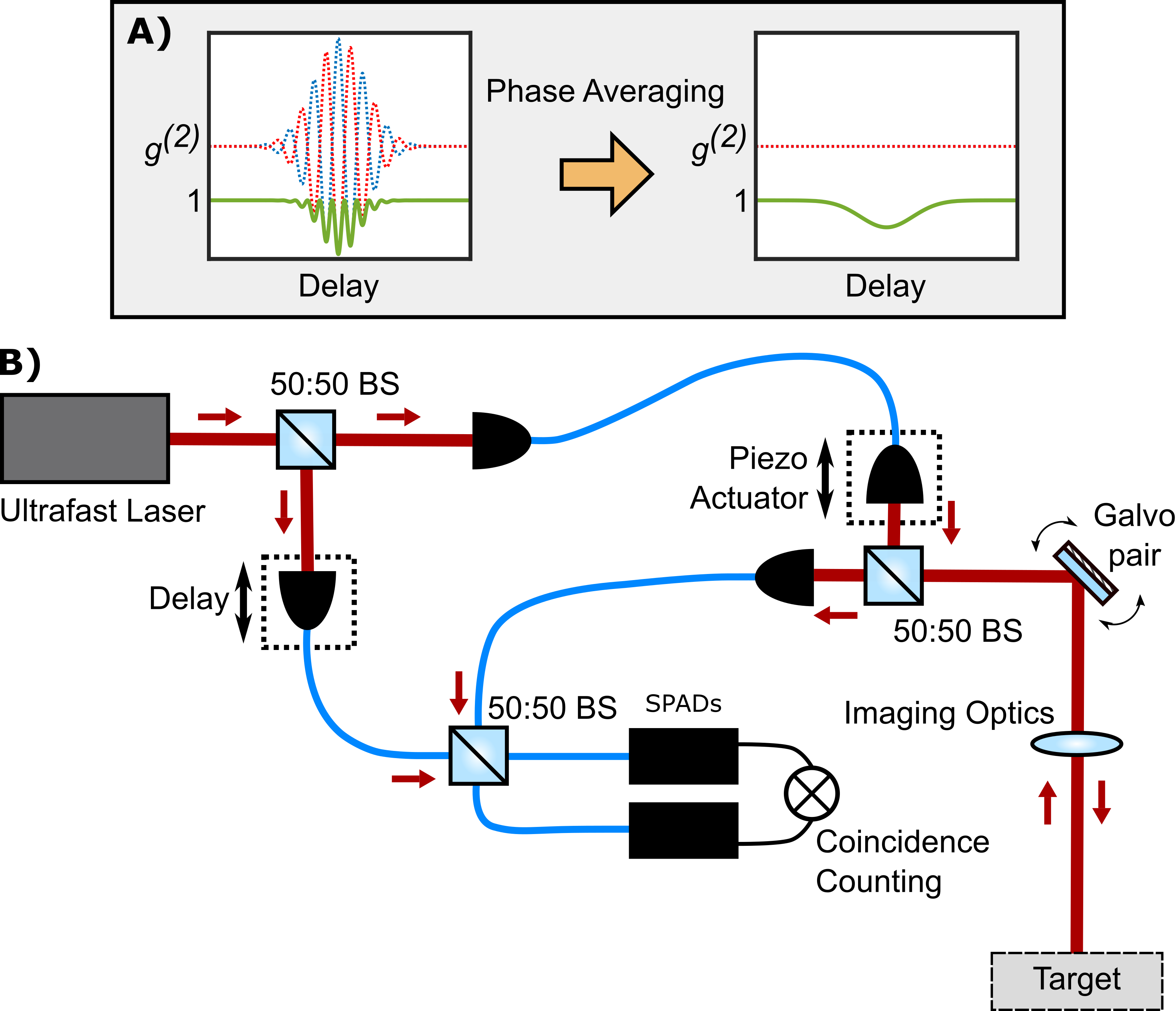}
\caption{A) Illustration of the working principle for the Two-Photon Interference LiDAR concept. The second order correlation function, $g^{(2)}$, from the two output ports of an interferometer are measured (green). Intensity of the two ports are also shown (red \& blue dashed). When there is a large amount of phase noise (Phase Averaging) the envelope function remains in the $g^{(2)}$. B) Schematic diagram of the experimental setup. Red arrows indicate beam propagation direction. From the first beamsplitter (50:50 BS), one arm of the interferometer (top left) is sent to the scene through a scanning mirror pair. The returned light is collected with a single mode fiber. The other arm (bottom right) is collected with a fiber with control over the optical path length (Delay). The phase of the interferometer is continuously randomised by a piezo actuator in one of the two arms. The two arms meet at a beamsplitter and where the output ports are monitored by a pair of SPADs.}
\label{fig_setup}
\end{figure}
Optical Coherence Tomography (OCT) is a 3D imaging approach whereby depth information is obtained from the interference of short coherence length light sources \cite{Huang1991a}. It is employed in a wide variety of bio-imaging applications perhaps most notably in ophthalmology, but also including and not limited to cardiology, oncology, and dermatology \cite{Fujimoto2000,Welzel2001,Zysk2007,Yonetsu2013,Shu2017}, owing to its ability to image within scattering media. Each interface inside a medium, usually a biological tissue, creates a reflection which interferes with a reference field. The location of the envelope function for each of these interference maxima is used to discern the distance light has propagated, or equally the Time-of-Flight (ToF), and thus the depth within the medium. In this way axial resolutions on the micron scale are routinely achieved with less than 1 $\mu$m having been demonstrated \cite{Bousi2010,Lichtenegger2017}.\\
OCT is dependent on the observation of the interference fringes arising from the cyclic phase of the optical field (regardless of whether they are observed in the temporal or spectral domain) and is thus an inherently coherent process. It is therefore not readily extended to incoherent systems e.g. situations where there is a mechanical movement on the wavelength scale or greater over the course of the acquisition time. These circumstances are abundant in LiDAR (Light Detection and Ranging) where interferometric stability cannot be achieved and even fluctuations in air currents can cause a substantial phase distortion. \red{Conventional single photon LiDAR is typically limited by the timing jitter of the detector which is on the order of tens of picoseconds or greater for SPADs \cite{Sanzaro2018}, resolving depth profiles on the millimetre scale and below therefore proves to be challenging.}\\
Hong-Ou-Mandel interferometry \cite{Hong1987} presents similar capabilities and the same depth resolution but in a way which is incoherent where we define ``incoherent" as possessing no first order coherence i.e. no dependence on the optical phase. For this reason, schemes based on HOM interferometry have been referred to as ``Quantum OCT" (QOCT) in the literature \cite{Abouraddy2002,Nasr2009}. QOCT instead gains ToF information by examining photon bunching at the output ports of an interferometer, usually by measuring correlated events between two single photon detectors, and mapping the envelope function directly without the need for phase dependent fringes. Recent works have shown how depth resolutions down to the sub-nanometre scale with HOM interferometry whilst also bypassing the need for any scanning elements \cite{Lyons2017}.\\
QOCT instead faces another challenge, HOM interferometry uses precisely 1 photon in each arm of the interferometer: \red{one which is sent to the scene and the back-scattered light collected that we refer to as ``probe" and the other kept in a controlled delay line that we refer to as ``reference"}, this is usually achieved with Spontaneous Parametric Down Conversion (SPDC). These sources a limited to at most 1 photon pair per pump laser pulse to remain in a photon starved regime and thus maintain high interference visibility. Where there are significant losses, this proves detrimental and cannot be mitigated by e.g. increasing the number of photon pairs without also introducing other negative effects such as limiting the non-ambiguous range of a LiDAR system. By way of example, consider SPDC photon pairs being generated at a rate of 10MHz with a loss rate of \red{$10^6$ (60 dB)}, only 10 coincident events per second would remain detectable. The generation rate could be increased to e.g. 1 GHz allowing 1,000 events to be measured per second but the range would be then limited to 30 cm. An equivalent of QOCT using classical states of light has also been demonstrated in the form of Chirped Pulse Interferometry (CPI) where benefits such as dispersion cancellation can be achieved without relying on entangled photon pairs \cite{Lavoie2009}. Interference between pairs of frequency combs has been used to achieve a similar goal, however requiring the use of a pair of femtosecond laser systems \cite{Wright2021}. \red{In addition to QOCT, there are a range of alternative schemes that combine interferometry with photon pair correlation measurements \cite{RadoslawChrapkiewicz2016,Defienne2021}.}\\
In the following, we explore the potential for using two-photon interference between weak coherent states of light specifically for LiDAR applications where it can provide the micron-scale depth resolution of OCT whilst also being immune to phase noise and able to operate in high loss environments. As well as demonstrating imaging capabilities for the first time, we investigate limitations on the achievable depth resolution and propose a new method for increasing the number of photon correlation resources to decrease the acquisition time. In the final sections of the article we discuss routes towards building a real-time imaging scheme based on this principle.

\section{Experimental Layout}

To demonstrate the concept, we construct a system (depicted in \cref{fig_setup}B) using a Ti:Saph oscillator (Coherent Chameleon Ultra II) with pulse duration of  130 fs at a rate of 80 MHz and a centre wavelength of 810 nm. This is split into two arms of roughly equal intensity to act as the reference and probe arms of the interferometer. The reference arm is coupled into a polarisation maintaining single mode fiber mounted on a translation stage to control the optical delay. The probe arm is also coupled into fiber before being directed onto a pair of scanning mirrors for transverse image scanning and the phase is randomised using a piezo actuator. The returned light is collected with the same optics and coupled back into single mode fiber before being combined with the reference arm at a fiber coupled beamsplitting cube. The use of single mode fibers (and a fiber-coupled beamsplitter) maximises the spatial mode overlap of two interferometer arms at the final beamsplitter and therefore the interference visibility. The photon number from each of the interferometer arms is further controlled and balanced using neutral density filters. The two output ports of the interferometer are measured using single pixel SPADs and the number of correlated photon pair events is measured using a TCSPC module (ID Quantique ID801). To demonstrate our system's high depth resolution, we acquire a depth scan of the two reflective surfaces of a 2mm thick glass diffuser (see Supplementary Material). Here, a deeply sub-mm depth response can readily be observed.\\

\section{High Resolution 3D Imaging}
\begin{figure}[b]
\centering
\includegraphics[width=0.7\textwidth]{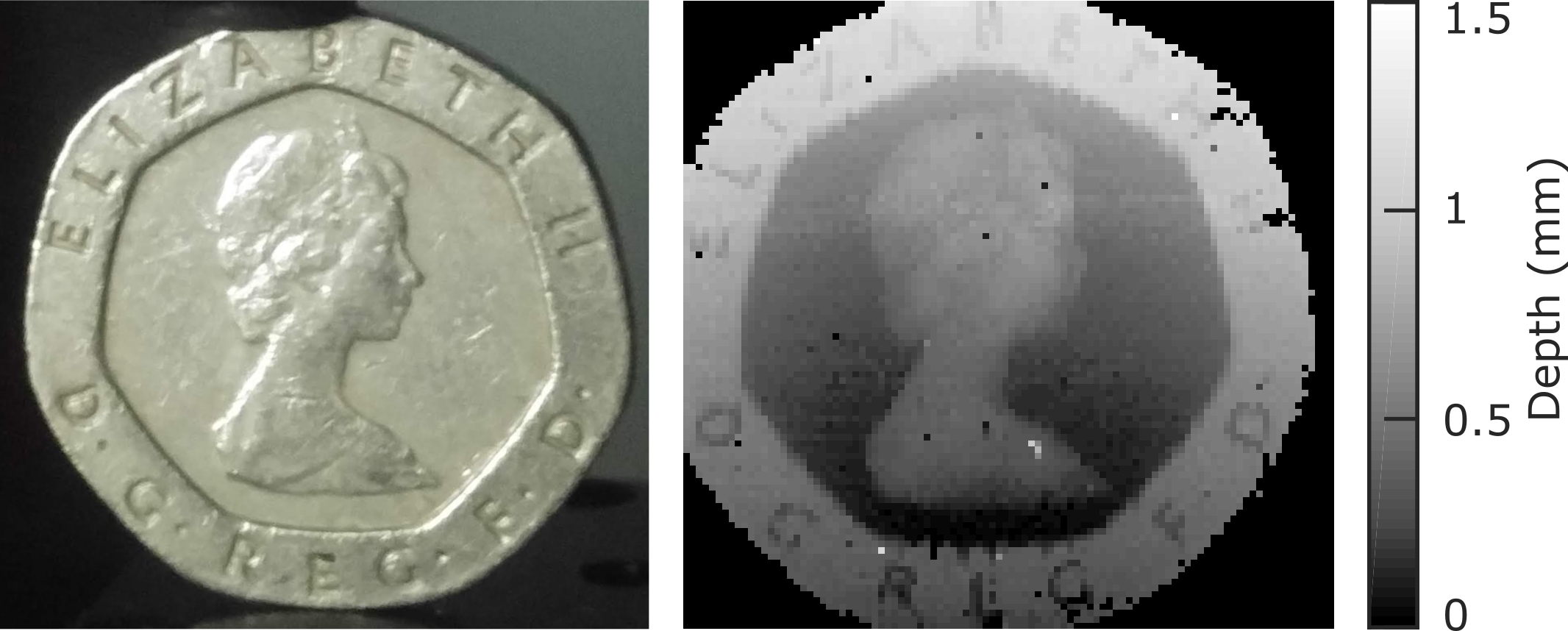}
\caption{Experimental demonstration of imaging two-photon interference LiDAR. Photograph of sample (left) and measured depth map (right) of a 20 pence coin. Estimated depth resolution is $7$ $\mu$m.}
\label{fig_coin}
\end{figure}
Full 3D imaging is performed by scanning the beam in the transverse plane using the galvanometer mirror system. Collimation optics are fixed in front of the illumination and collection fibers and the far-field of the scanning mirrors is imaged onto the scene, positioned 1 m away, by a single lens. The scene is scanned in a grid of 100 $\times$ 100 positions and the depth is estimated by taking the center position of a Gaussian function fitted to each delay scan, details of which can be found in the Supplementary Material. \red{We find that the depth resolution for our system, estimated from the uncertainty in the $g^{(2)}$ minimum position from our fits, to be approximately 7 $\mu$m.}\\
The resulting depth map for a scene consisting of a 20 pence coin is shown in \cref{fig_coin}. We can determine an effective IRF for our system from the width of the correlation dip where, in the specific case of \cref{fig_coin}, we measure this to be 70.4 $\mu$m. The depth scan was achieved using 75 steps with a 50 ms integration time per point, giving a total acquisition time for the image of approximately 10 hours.\\
\begin{figure}
\centering
\includegraphics[width=0.7\textwidth]{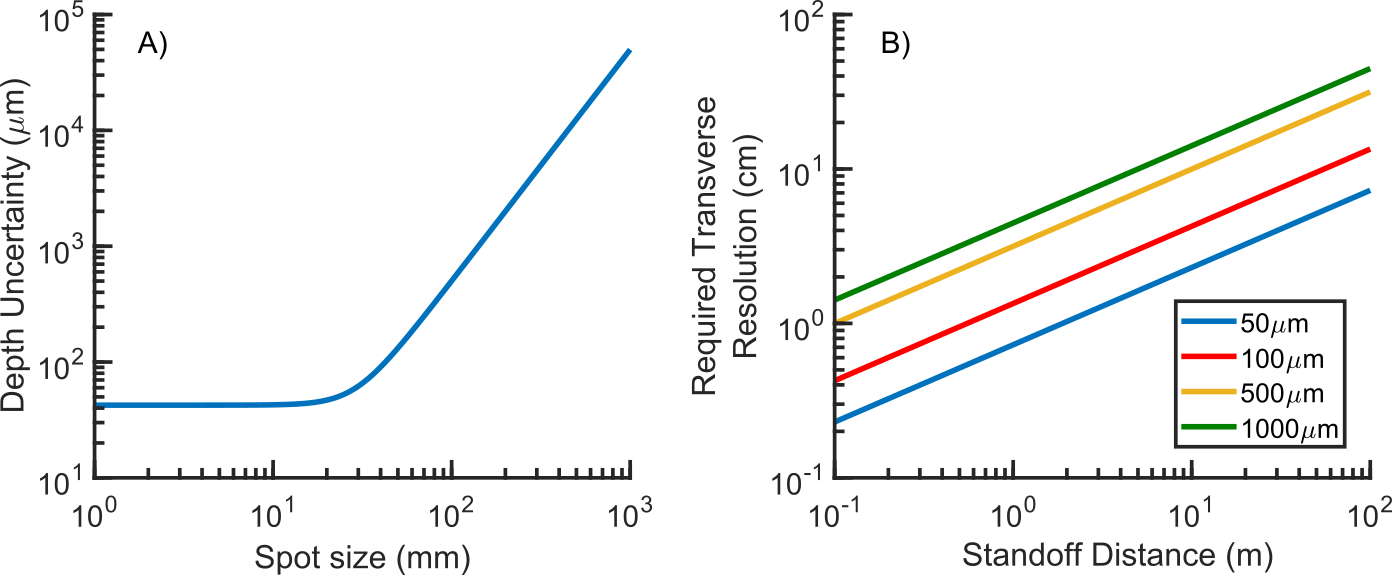}
\caption{A) Numerical estimation of the axial resolution as a function of transverse resolution for a fixed standoff distance of 10 m and pulse duration 100 fs. B) The minimum required transverse resolution to achieve a given depth resolution for the same 100 fs pulse.}
\label{fig_res}
\end{figure}
Due to the high depth resolution (or equivalently, time-of-flight resolution) this method is capable of, and the relatively large distances of interest to LiDAR, small path length differences owing to the geometry of the scene must be taken into account as they produce not only a transversely dependent offset to the photon time of arrival but will also effect the axial resolution. At a given stand-off distance from the scene, the spot size of the transverse scanning system (or equally the transverse resolution for a full-field system) acts as an additional source of axial uncertainty with larger spots producing a greater spread of arrival times. At a given distance, $Z$, transverse spot radius, $r_\text{spot}$, and pulse duration, $\tau_\text{pulse}$, the depth uncertainty can be approximated as
\begin{equation}
\Delta Z = \sqrt{\left(\sqrt{Z^2 + r_\text{spot}^2} - Z \right)^2 + 2\tau_\text{pulse}^2}.
\end{equation}
This effect is numerically estimated in \cref{fig_res}A for a stand-off distance of 10 m and a pulse duration of 100 fs. As expected, for small spot sizes, the axial (depth) resolution is limited by the duration of the optical pulse.  However, for beam diameters greater than 10 mm the geometric uncertainty becomes visible and scales as $r_\text{spot}^2$. The impact of the transverse spot size on the axial resolution reduces as the distance to the scene is increased: \cref{fig_res}B shows the required transverse resolution as a function of the standoff distance to achieve axial resolutions of 50, 100, 500, and 1000 $\mu$m. This is due to a parallax effect whereby the difference in path length from the sensor to one transverse extreme point in the scene, and the sensor at the other extreme, decreases.\\

\section{Photon Number and Loss}
We evaluate the losses and compare to QOCT and other methods that rely on correlation between photon pairs from e.g. SPDC. Here we assume a loss that scales linearly with propagation distance as will occur in schemes where only one of the two photons is sent to the scene. Using a diffuser (see SM) as an example of realistic backscattering conditions (FWHM of scattering cone $\approx$ 16$^\circ$ from specification) and a distance of 100 m, a 2 inch optic would only be able to collect approximately 2$\times 10^{-4}$ \% of the returned light \red{(56 dB loss)}. An SPDC source generating on the order of $10^6$ photon pairs per second would therefore only be able to measure 1-2 correlated events within the time used to acquire a single temporal scan from the experiment shown here (3.75 s). Where, with the proposed Two-Photon Interference LiDAR method this loss is compensated by simply increasing the laser power, generating more photon pairs via optical nonlinearities without also introducing substantial numbers of multi-pair events remains a significant technical challenge.\\
We note that, due to the use of coherent states as opposed to photon pairs from SPDC, correlated events need not be limited to photons from the same optical pulse. Consider a scene that is imaged using pulses at a repetition rate on the MHz scale that only exhibits phase noise on the kHz scale, pulses within a 1$\mu$s window of each other observe the same interferometer phase and therefore the same behaviour will be seen in the $g^{(2)}$ regardless of whether correlations are taken between pairs of photons from different pulses within this timescale or the same pulse. To demonstrate how this can be exploited we conduct an additional measurement where we record the individual timetags of each detected photon and examine correlations between neighbouring pulses up to a decorrelation time $\tau_\text{corr}$. The total number of correlated events in this time can then be summed resulting in a vast increase in the total number of photon counting resources from the same physical measurement, thereby also allowing the measurement time to be decreased whilst maintaining the same SNR. We find that, for our specific system, up to the closest 1000 pulses can be used within the time $\tau_\text{corr}$ allowing for an acquisition time three orders of magnitude faster. A comparison of a correlation dip with and without our method can be seen in \cref{fig_sum_corr}. Full details of this approach can be found in the Supplementary Materials.
\begin{figure}[hb]
\centering
\includegraphics[width=\textwidth]{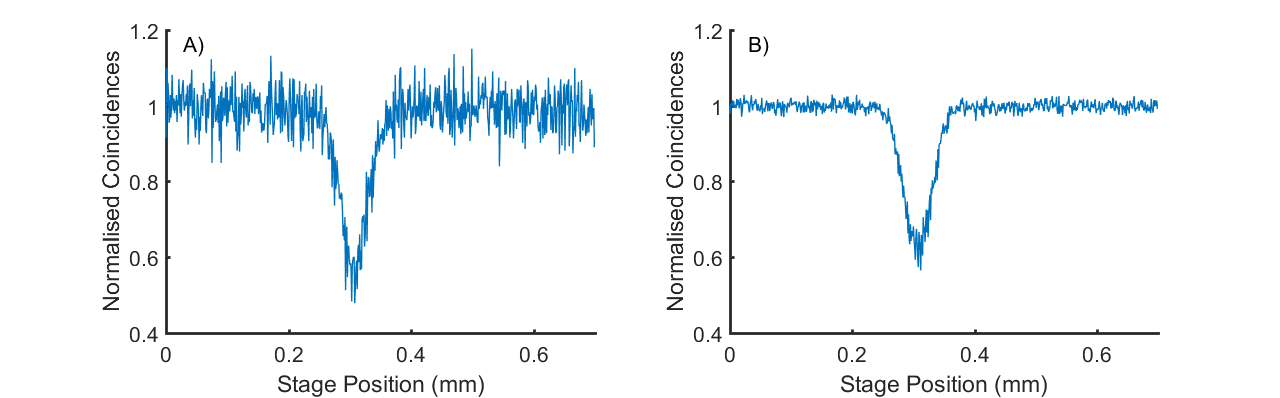}
\caption{Improvement in the noise of the correlation dip using our approach. A) The standard approach to measuring correlated events where a coincidence is registered when a photon arrives at both detectors on the same pulse. B) The dip recovered using our method looking at correlations between, in this case, the closest 1000 pulses.}
\label{fig_sum_corr}
\end{figure}

\section{Discussion}
\red{In our measurements, we use prior knowledge about the object position to limit the scan length and therefore overall measurement time. We note, however, that the approximate position of objects of interest in the scene can be instead gained using conventional single photon LiDAR as all the same components are present here (pulsed laser, single photon detector, TCSPC).} The axial resolution demonstrated here can be increased further by the use of pulses of shorter duration. It then also becomes important to consider the effects of chromatic dispersion from propagation within the air and fibers which will broaden the pulse. We note, however, that the 100 fs pulse discussed here will only broaden by around 14 fs for 100 m propagation in air due to second order dispersion \cite{Walmsley2001,Wrzesinski2011}. This can be compensated within either the signal or reference arm and could be optimised without any prior knowledge of the propagation distance.
Although the measurement shown in \cref{fig_coin} required over 10 hours to acquire, this can be greatly reduced by using our proposed multiple-pulse correlation method to just 36 s. Further improvements can be achieved by the removal of the transverse scanning elements in favour of a full-field imaging system. Full-field OCT is routinely performed and it is only the addition of the correlation measurements that would be required to extend this to the approach presented here. To achieve this, cameras which are able to distinguish $g^{(2)}$ from individual photon correlations at a high rate is preferable. Recent advancements using SPAD cameras have proven to be a promising candidate for this \cite{Ndagano2020}. \red{Full-field imaging with HOM interference has been recently demonstrated using SPAD cameras, the scheme however relied on the spatial correlation between photon pairs from an SPDC source \cite{Ndagano2022}. The same method can be adapted for our LiDAR scheme by using spatially correlated light sources.} \\

\section{Conclusions}
We demonstrate a new method for LiDAR using intensity correlation based interferometry with the depth resolution of an OCT system. We have shown imaging capabilities with the ability to discern two different depth objects at the same pixel down to separations of 70 $\mu$m. Our approach has applications not only for high-precision LiDAR, but also in other ToF sensing systems where one wishes to measure the full temporal distribution of light pulses with high resolution such as imaging through scattering media and Non-Line-of-Sight imaging.\\
\\
\section*{Acknowledgements}
\red{The authors acknowledge support from the Royal Academy of Engineering on grant RF\textbackslash202021\textbackslash20\textbackslash329 and EPSRC on grant EP/R030081/1.}\\
\\
\textbf{Disclosures.} The authors declare no conflicts of interest.\\
\red{\textbf{Data Availability.} Data used to produce all figures shown in this paper can be found at \url{http://dx.doi.org/10.5525/gla.researchdata.1306}}\\

\noindent See Supplement 1 for supporting content

\bibliography{opt_coh_lidar_ref}






\end{document}


\title{Supplementary Material: Two-Photon Interference LiDAR Imaging}
\author{Robbie Murray}
\affiliation{School of Physics and Astronomy, University of Glasgow, Glasgow, G12 8QQ, UK}
\author{Ashley Lyons}
\affiliation{School of Physics and Astronomy, University of Glasgow, Glasgow, G12 8QQ, UK}
\maketitle
\section{Principle of Operation}

The scheme consists of a Mach-Zehnder style interferometer (as depicted in fig.1 of the main text) with a pulsed laser source. The probe arm is directed to the LiDAR scene of interest whilst an appropriate delay is added to the reference. The second order coherence function, $g^{(2)}$, is calculated by evaluating coincident events between two single photon detectors after the final beamsplitter. The interferometer is operated in an incoherent regime by randomising the phase of one arm and the envelope function manifests as a dip in the $g^{(2)}$, the location of which reveals the group delay of the probe pulse. In this way, the system is the coherent state equivalent to a HOM interferometer where the $g^{(2)}$ is limited to a 50\% reduction \cite{Rarity2005,Kim2014}.
The system can be compared to a conventional single photon LiDAR scheme which typically consists of a pulsed laser source and a single photon detector plus TCSPC electronics to resolve the time of arrival. In the conventional LiDAR case, the ability to resolve two or more objects closely positioned in depth at the same transverse position is limited by the IRF of the detector and timing electronics. This is on the scale of $>$30 ps for Single Photon Avalanche Diodes (SPADs) \cite{Sanzaro2018}, equating to a free-space propagation distance of around 9 mm. Conversely, the temporal IRF of our system is determined by the optical pulse duration rather than the detection electronics and can therefore readily reach femtosecond timescales, or a depth separation on the order of microns. We note, however, that the detectors still require a high enough bandwidth to resolve pulse-to-pulse correlations which will be on the nanosecond scale for MHz repetition rates, this is a much looser constraint than conventional LiDAR on the picosecond timescale where the detectors provide the temporal information directly. We also note that the use of single photon detectors is not a requirement here, and that higher intensities could be used with e.g. photodiodes provided the bandwidth criterion is met.

\section{Ranging of multiple surfaces.}
\begin{figure}[b]
\centering
\includegraphics[width=0.6\textwidth]{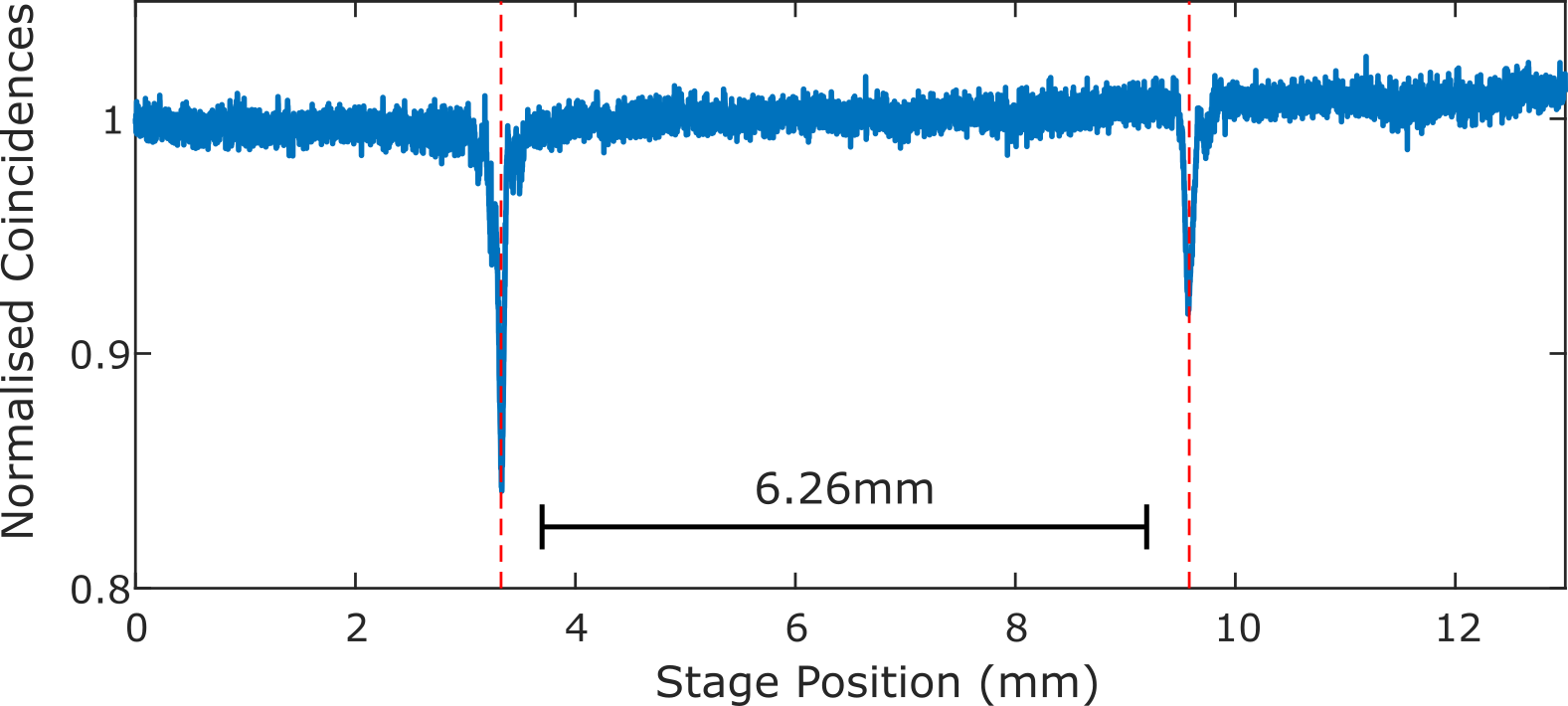}
\caption{Depth profile of the light reflected from a glass diffuser. The time-of-flight signal returns a dip in the photon coincidences from the first (left) and rear (right) surface of the glass.}
\label{fig_diff}
\end{figure}

The system's ability to perform depth ranging  is assessed by using a 2 mm thick ground-glass plate as the target that has a smooth surface placed furthest away from the transceiver and a rough, diffusing surface closest to the transceiver. For this test no lenses were used for the illumination and collection fibers giving a beam radius of approximately 10 mm at a distance of approximately 10 cm between the glass plate and the illumination/collection fibers. The results are shown in \cref{fig_diff}. Two dips are observed in the correlation trace of the returned signal: the first due to the first scattering surface, and the second is due to the back reflection from the smooth rear surface. To validate the depth profiling result, the thickness of the diffuser is estimated from the difference in the time of flight of the two observed dips. The centres of the two dips is estimated from a Gaussian fit applied to each dip, resulting in a difference in position of 6.26 mm (20.1 ps) which, also accounting for the double-pass through N-BK7 glass and taking a group index of 1.53 from literature \cite{schott_cat}, results in a thickness of 2.04 mm. This is well within the manufacturer specified value of 2 $\pm$ 0.2 mm. We note that the IRF of the SPADs used here is 350 ps implying that we would be unable to resolve the difference between these two return signals using the conventional LiDAR TCSPC-based approach. This observation indeed, encompasses the key aspect of this approach - the temporal and hence depth resolution depends on the photon or laser pulse spectral bandwidth and not on the detector IRF.\\

\section{3D Imaging Fitting Procedure}
For each transverse position that is scanned to produce a 3D image (fig. 2 in main paper), a Gaussian fit is performed to determine the estimated depth. Transverse positions that yield a poor depth estimation are identified using the returned R squared value from the fit and subsequently removed. This is done to remove pixels falling outside the object as well as pixels where the highly specular reflection from the coin caused saturation of the SPAD detectors. A linear interpolation across adjacent image points is then used to fill in these missing pixels. Despite the small size of the object, the sub-mm depth features can clearly be resolved including the writing around the edges, and the contours of the face and crown. A slight slant in the image is visible due to the coin being positioned slightly above the imaging optics. Each depth scan was acquired using 75 points over a range of 1.5 mm with an integration time of 50 ms per point. This equates to a total acquisition time of over 10 hours. However it is noted that by performing the same measurement with a full-field imaging system, as is common in conventional OCT and recently been demonstrated for HOM \cite{Ndagano2022}, this can be reduced to under 4 s. \red{It should be noted, however, that previous full-field imaging using HOM required carefully matching the spatial mode size to the pixel pitch.} The scanning in the time domain (delay) can also be circumvented by performing the measurements in the spectral domain, in much the same way as OCT is typically performed. It has already been demonstrated that QOCT can be extended to the spectral domain, including using classical light sources\cite{Olenderska2008,Lavoie2009}.\\
The average uncertainty in the depth estimation, calculated from the confidence bounds of each fit, was estimated to be $\pm7$ $\mu$m. For applications where the two or more objects are positioned closely in depth, it is the width of the temporal IRF that is of the main concern. In our case, this is given by the width of the correlation dip which had an average standard deviation of the dips for the data shown in fig.2 of the main text is 70.4 $\mu$m. The resolution is ultimately limited by the pulse duration and can be estimated from the width of the autocorrelation. Assuming a Gaussian pulse, this is estimated to be approximately 55 $\mu$m. The discrepancy between this and the measured value is most likely due to experimental noise.\\
\begin{figure}[b]
\centering
\includegraphics[width=0.6\textwidth]{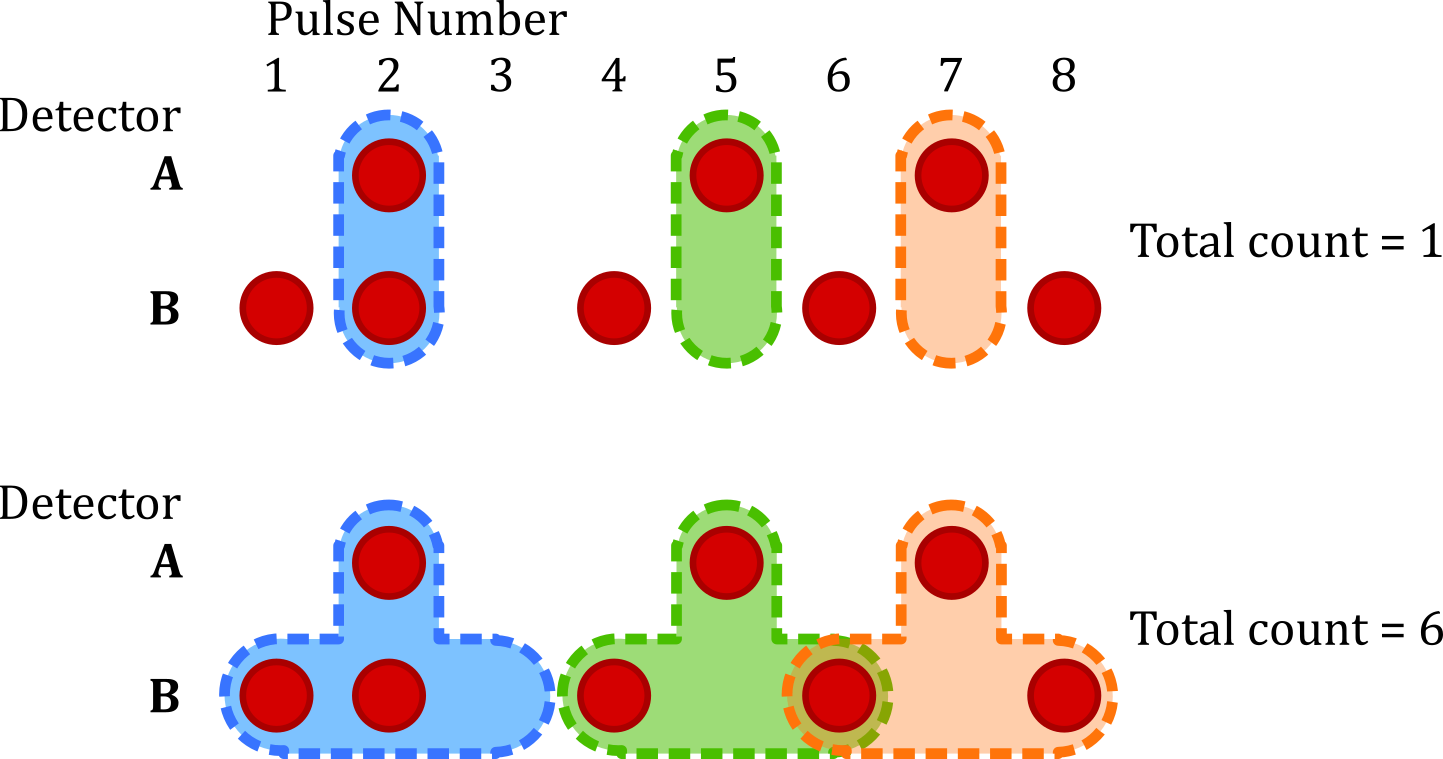}
\caption{Illustration of the coincidence counting scheme, red circles indicate a photon detection, coloured bubbles show the coincidence ``window". (Top) The typical method for counting photon-photon correlations, a coincidence is counted when a photon is observed at ``B" conditional on a photon being observed at ``A". (Bottom) Our scheme, using correlations over a finite-sized window, here chosen to be 3 pulses. The coincidence event number is equivalent to the total number of photons detected at ``B" within the window conditional on a photon at ``A". Photons at ``B" may be counted more than once, such as in pulse 6, but correspond to statistically independent measurements. In this example, each window contributes 2 to the total coincidence count.}
\label{fig_corr_scheme}
\end{figure}

\section{Increasing Correlation Resources}
In the following we detail how correlated photon events across multiple laser pulses may be used as additional photon counting resources, thereby achieving the same Signal to Noise \red{Ratio} (SNR) at a reduced acquisition time.\\
\\
Suppose de-phasing in the LiDAR scheme occurs due to mechanical motion of an object within the scene. Within a certain amount of time, $\tau_\text{corr}$, the scene will appear static and the interferometer shown in Fig. 1 will remain in-phase, where $\tau_\text{corr}$ is dependent on the velocity of the objects within the scene. If the sampling rate of the scene (laser repetition rate) is sufficiently high then multiple laser pulses will observe the same interferometer phase up to the time $\tau_\text{corr}$. There then exists scenarios where one may take correlated photon events where one of the photons is measured from one pulse and the other photon is measured from a subsequent pulse which have the same result as if the photons both originated from the same laser pulse i.e. everything occurs with the same interferometer phase. These cross-pulse correlations may be used as an additional resource to decrease Poisson noise from the same amount of data. We note that we explore the case here for a phase-randomised coherent state and the same is not necessarily true for e.g. entangled photon pairs generated via Spontaneous Parametric Down Conversion (SPDC).\\
\begin{figure}[b]
\centering
\includegraphics[width=0.6\textwidth]{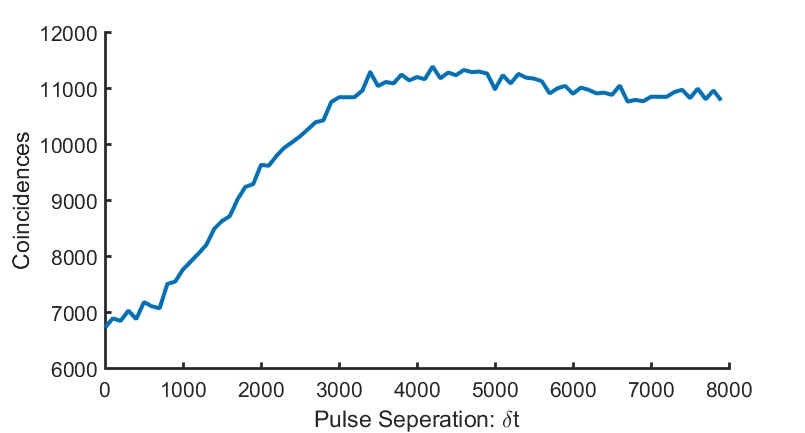}
\caption{Measurement of $\tau_\text{corr}$ for our system. The stage position is set to the bottom of the HOM dip and the number of correlated events between time delayed pulses is determined. Up to a separation of approximately 1000 pulses, or 12.5 $\mu$s at an 80 MHz rep rate, the photons remain coherent and a clear reduction in the correlated events can be observed due to the HOM effect.}
\label{fig_tau_corr}
\end{figure}
\\
Our proposed method is outlined in \cref{fig_corr_scheme}. Ordinarily when examining photon-photon correlations (e.g. when using entangled photon pairs) with a pulsed laser source, a ``coincidence" event is counted only when a photon is detected at both detectors (which we label ``A" and ``B") at the same pulse. However, as discussed above, one can also take a coincidence event between a detected photon at ``A" and a second photon from any arbitrary pulse at ``B" within the time $\tau_\text{corr}$ and obtain equivalent results. It is therefore possible to sum the number of coincidence events across \emph{all} detected photons at ``B" within $\tau_\text{corr}$ for a photon detected at ``A" to greatly increase the total event number. This process can then be repeated for all detected photons at ``A". Although the individual photons contribute to multiple pair-wise coincidence events, each coincident event remains random and independent from one another and thus able to increase the SNR by reducing the Poissonian uncertainty.\\
\\
We first identify $\tau_\text{corr}$ to establish the magnitude of the potential benefit for a physical system. In the following we remove the target object and couple the signal arm of our interferometer straight into fiber to achieve higher visibility interference. The optical delay is fixed at the bottom of the HOM dip (highest visibility interference) and keep a record of all photon timetags from the TCSPC module (Picoquant Hydraharp). We then measure the number of coincidence events between two temporally separated pulses, i.e. without the summing described above, as a function of the separation, $\delta t$. This is shown in \cref{fig_tau_corr}, \red{at $\delta t$ close to zero the interferometer remains coherent and the reduced coincidence events, corresponding to the  minimum of the HOM dip,} is observed as expected. As $\delta t$ is increased coherence is lost and the visibility is reduced up to the ``baseline" number of coincident events ($g^{(2)} = 1$). As can be seen, for our system $\delta t <$ 1000 pulses produces no significant change in the visibility. We use this as the maximum range for our ``summed" coincidence events in the following. \red{We can also use this information to estimate the piezo dither rate in our setup, which will be the inverse of $\tau_\text{corr}$. We fit \cref{fig_tau_corr} with a Gaussian function and find the FWHM to be approximately 3200 pulses equivalent to 40 $\mu$s at our repetition rate. The dither is therefore around 25 kHz.} \\
\\
To illustrate the benefit of this approach, we show the HOM dip measured with only same-pulse coincidences in fig. 4A and coincidences summed across the nearest 1000 pulses in fig. 4B for the same acquisition time of 0.25 ms per stage position. We note that, where dominated by Poisson noise, the improvement to the SNR will be equal to $\sqrt{n_\text{corr}}$, where $n_\text{corr}$ is the number of pulses used, however the impact to the total experiment time is $n_\text{corr}$. Or, in other words, our approach recovers $n_\text{corr}$ times more resources in the same amount of time thereby allowing the measurement time to be reduced by $n_\text{corr}$ whilst achieving the same SNR.
\section{Signal to Noise Ratio of Coin Measurement}
\begin{figure}[t]
\centering
\includegraphics[width=0.6\textwidth]{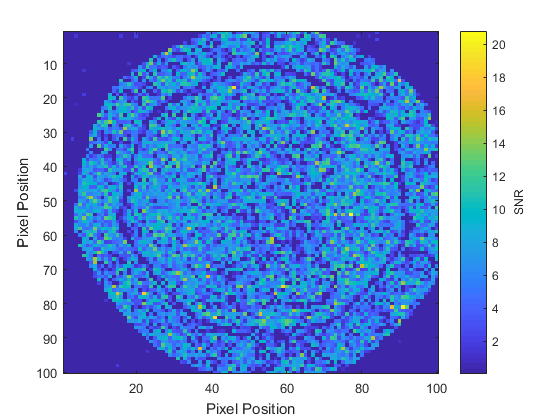}
\caption{Measurement of the SNR for each HOM dip used to produce Figure 2 in the main text.}
\label{fig_coin_snr}
\end{figure}

\red{Here we quantify the SNR of the individual HOM dips used to produce our coin depth map (Figure 2 in the main text). Firstly, we define SNR in our case to be the amplitude of the HOM dip (signal) divided by the standard deviation of data points where there is no interference i.e. outside of the HOM dip (noise). For the signal, we take the amplitude of the Gaussian fit described in Section III. To identify positions in the delay scan outside of the HOM dip to characterise the noise, we perform a circular shift of each of the HOM dips to systematically position the minimum close to the beginning of the scan range. We then take the standard deviation of the final 35 points to ensure we isolate a region with no interference. The resulting SNRs are shown in \cref{fig_coin_snr} using the same thresholding to filter out pixels where the data quality is judged to be too poor. There is a large variation in the SNR due to the specular reflection from the metallic surface causing a large change in the measured photon counts.}
\bibliographystyle{unsrt}
\bibliography{opt_coh_lidar_ref}